\documentclass[%
 aps,prb,
 amsmath,amssymb,
 showpacs,reprint]{revtex4-1}

\usepackage{graphicx}
\usepackage{dcolumn}
\usepackage{bm}

\usepackage[latin1]{inputenc}
\usepackage[T1]{fontenc}
\usepackage{mathptmx}
\usepackage{gensymb}

\newcommand{\UGA}{Universit\'e Grenoble-Alpes}
\newcommand{\npsc}{CEA-CNRS joint group ``Nano-Physique \& Semi-Conducteurs''}
\newcommand{\pheliqs}{CEA, IRIG, PHELIQS, 17 rue des martyrs, 38054 Grenoble {\sc cedex} 9, France}
\newcommand{\Neel}{CNRS, Institut N\'eel, 25 rue des martyrs, 38042 Grenoble {\sc cedex} 9, France}
\newcommand{\aztse}{Ag$_2$ZnSnSe$_4$}
\newcommand{\gap}{\text{$E_\text{g}$}}
\newcommand{\voc}{$V_\text{OC}$}
\newcommand{\fermi}{\text{$E_\text{F}$}}
\newcommand{\boltzmann}{\text{$k_\text{B}T$}}

\begin{document}

\title[Sample title]{Optical Determination of the Band Gap and Band Tail \\
of Epitaxial \aztse\, at Low Temperature}

\author{S. Perret}
\affiliation{\UGA, \npsc, \pheliqs}%
\author{Y. Cur\'e}
\affiliation{\UGA, \npsc, \pheliqs}%
\author{L. Grenet}
\affiliation{\UGA, CEA, LITEN, 17 rue des Martyrs, 38054 Grenoble {\sc cedex}, France}%
\author{R. Andr\'e}%
\affiliation{\UGA, \npsc, \Neel}%
\author{H. Mariette}%
\affiliation{\UGA, \npsc, \Neel}%

\author{J. Bleuse}
\email{joel.bleuse@cea.fr}
\affiliation{\UGA, \npsc, \pheliqs}%

\date{\today}

\begin{abstract}
We report on the precise determination of both the band gap \gap\, and the
characteristic energy $U$ of the band tail of localized defect states, for
monocrystalline \aztse. Both photoluminescence excitation and time-resolved
photoluminescence studies lead to $\gap = 1223\pm3$~meV, and $U = 20\pm3$~meV,
at 6~K. The interest of the methodology developed here is to account
quantitatively for the time-resolved photoluminescence and photoluminescence
excitation spectra by only considering standard textbook density of states,
and state filling effects. Such an approach is different from the one most
often used to evaluate the energy extent of the localized states, namely by
measuring the energy shift between the photoluminescence emission and the
excitation one --- the so-called Stokes shift. The advantage of the present
method is that no arbitrary choice of the low power excitation has to be done
to select the photoluminescence emission spectrum and its peak energy.
\end{abstract}


\maketitle

\section{Introduction}

The occurrence of a band tail just below the band gap in the absorption
spectra, when adding impurities to  a pure crystal was first reported a long
time ago by F. Urbach \emph{et al.} \cite{Urbach1953_PR92,*Moser1956_PR102} in
AgCl crystals containing copper as an impurity. The observed band tail was
well described by a single exponential and attributed to localized states
created by the Cu-Ag cation disorder. Such analytic contribution to account
for optical properties of doped semiconductors was then found to be verified
and useful in a huge number of cases, especially when dealing with localized,
substitutional impurities. More recently, as an example, the optical signature
of impurity-impurity interactions in copper-containing II-VI alloy
semiconductors has been identified by systematically comparing optical
absorption and emission spectra \cite{Bhattacharyya2018_JPCL9}. We show here
such an experimental, optical approach; we quantitatively analyze
photoluminescence excitation data (reflecting the density of states) and
time-resolved emission spectra (reflecting the evolution of the electron-hole
population), using a single exponential shape for the localized states, as
proposed by Urbach. This allows us to accurately determine the band gap and
the energy extent of the band tail in the case of \aztse, a new promising 
material for solar energy conversion.

Indeed, by contrast to solar cells based on chalcopyrites such as
Cu(In,Ga)(S,Se)$_2$ and zinc blende structures (CdTe), for which a power
conversion efficiency (PCE) above 20\% has been achieved
\cite{Green2019_PP27}, k\"esterite-based solar cells such as
Cu$_2$ZnSn(S,Se)$_4$ (CZTSSe) only reached 12.6\% PCE \cite{Wang2014_AEM4}.
This limited performance is mainly related to a large open-circuit voltage
(\voc) deficit, as evidenced by the above CZTSSe record cell, which attains
only 59\% of the Shockley-Queisser limit for \voc, whereas it reaches 80\% of
this limit for both the short circuit current and the fill factor
\cite{Wang2014_AEM4}.

Several hypotheses have been drawn to explain such a \voc\, limitation. Among
them, the presence of a large band tail of localized states below the extended
state bands is considered to be one of the main culprits, as it diminishes the
effective bandgap, and consequently the available \voc\,
\cite{Gokmen2013_APL103}; it also induces the localization of charge carriers,
and thereby reduces their collection efficiency in devices.

The presence of large densities of intrinsic native point defects, as
predicted by Density Functional Theory (DFT) calculations
\cite{Chen2013_AM25,Chen2010_PRB81,Chen2012_APL101}, can directly influence
the generation, separation and recombination of photo-generated electron-hole
pairs. For CZTSSe, a large density of acceptor Cu$_\text{Zn}$ anti-sites and
Cu vacancies V$_\text{Cu}$ has been calculated, given that both these point
defects have the lowest formation energies \cite{Chen2013_AM25}.  Moreover,
these point defects are stabilized by the formation of self-compensated
neutral defect complexes made of triplets of exchange atoms such as
[2Cu$_\text{Zn}$ + Sn$_\text{Zn}$]. In other words, the disorder in the cation
sublattice, especially between Cu and Zn, strongly influences the occurrence
of such defect clusters \cite{Chen2010_PRB81,Chen2012_APL101}, and is then
directly responsible for the presence of potential fluctuations
\cite{Romero2011_PRB84}: large amounts of these defects are likely to form
band tailing, and believed to be one of the main reasons for the \voc\,
deficit \cite{Gokmen2013_APL103,Rey2018_SEMSC179}. This issue is of particular
importance, which is why we pursued our optical investigation in depth.

Such band tails have been studied qualitatively by optical spectroscopy:
mainly photoluminescence (PL) as a function of temperature and excitation
power \cite{Tanaka2006_PSSa203, Leitao2011_PRB84, Lin2015_APL106,
Oueslati2015_SEMSC134, Sendler2016_JAP120, Lang2017_PRB95,
Grossberg2012_APL101, *Grossberg2014_CAP14}, but also by PL excitation
spectroscopy (PLE), which reveals the density of states in the presence of
potential fluctuations \cite{Siebentritt_2005_PSSb242,Bleuse2018_JEM47}, and
by time-resolved photoluminescence spectroscopy (TR-PL), which gives
information about the localization and transfer of carriers between the band
tail states, as was done on CZTS single crystals
\cite{Lang2017_PRB95,Bleuse2018_JEM47}: all these data have been analyzed as a
function of the order/disorder degree in the quaternary structure
\cite{Bleuse2018_JEM47,Timmo2017_EMRS13}.

The substitution of copper by silver is one of the proposed methods to reduce
the density of defect states. The motivation for \aztse\, (AZTSe) or
(Cu$_{2-x}$Ag$_x$)ZnSnSe$_4$ derives from the reduced Ag$_\text{Zn}$
anti-sites concentration that was theoretically predicted by DFT calculations,
as a consequence of a much larger formation energy
\cite{Nakamura2011_PVSEC26,Yuan2015_AFM25,Chagarov2106_JCP144} for
Ag$_\text{Zn}$ than for Cu$_\text{Zn}$ anti-sites.  Then, if the anti-site
defects between atoms of column I and II of the periodic table are indeed the
main cause of band tailing, \aztse\, has the potential to significantly reduce
the problem and to allow for larger \voc.

Such a substitution was successfully used \cite{Cui2014_APL104,
Gershon2016_AEM6_10, Hages2016_SEMSC145} to observe an improved efficiency,
mainly due to a 8--10\% improvement of \voc\, as compared to the baseline
CZTSSe device. More precisely, it has been estimated, from room temperature
optical data, that the band tailing effect is dramatically suppressed for
AZTSe (pure-Ag) samples \cite{Gershon2016_AEM6_10}.

In this work, we study single crystal, stoechiometric \aztse\, epilayers, as
they offer a way to reduce the ambiguity on the nature of defects by
eliminating grain boundaries. Complementary, low temperature optical
experiments are compared : (i) the variation of the PLE intensity as a
function of excitation photon energy; (ii) the evolution in time and energy of
the TR-PL emission after a short pulse excitation; (iii) the variations of PL
spectra with excitation power density. All these data are analyzed with a
quantitative model that  enables us to accurately determine both the band gap
and the energy extent of the localized states band tail, which appears just
below the band gap. These experimental data also evidence the transfer
mechanisms that occur between the localized states, within this band tail.

\section{Experimental measurements}

The samples are grown in a MBE chamber using four Knudsen cells for Ag, Zn,
Sn, and Se, without post treatment of the layer. The beam equivalent pressures
of the four elements were measured by a gauge pressure meter positioned at the
location of the sample. The AZTSe layers are grown on a (001) InAs substrate
kept a temperature of 460~\celsius\, with a growth rate of 4~nm per minute.
X-ray diffraction and Raman spectroscopy are used to check the absence of
secondary phases, as was done on CZTSe \cite{Cure2017_SM130}.

For optical spectroscopy, both steady-state and time-resolved
micro-photoluminescence experiments are carried out at cryogenic temperatures
(6~K) in a helium-flow, optical cryostat. Electron-hole pair injection in the
AZTSe layer is provided by a Ti:sapphire laser (Coherent Mira), operating
either in steady-state mode or in pulsed mode with 200~fs-long pulses,
and a repetition rate set by a cavity dumper at either 246~kHz (extracting one
pulse every 222) or 501~kHz (every 109). The laser excitation is focused down
to a 1.5-$\mu$m diameter spot on the sample with the same microscope objective
(0.4 numerical aperture) that collects the luminescence signal. For PLE
experiments, the excitation source is a 1~kW, halogen lamp coupled to a
monochromator. The emission of the AZTSe layer is then spectrally dispersed by
a 640~mm-focal length monochromator and detected by a silicon CCD (Andor
DU420A-BRDD, for steady-state PL and PLE) or a silicon avalanche photodiode
(Perkin-Elmer SPCM AQR-15, for TR-PL).

Low-temperature TR-PL spectra are measured at a series of about 20 different
photon energies, evenly spaced from 1100 to 1370~meV, over a time window of
about 600~ns, with 50-ps time bins. This gives a full view over the relaxation
and recombination processes involved in the \aztse\, sample. For each of the
nearly 12000 time bins, a low-resolution PL spectrum is therefore available
for analysis. The temporal instrument response of the TR-PL setup, measured
with the excitation laser pulses, presents a time resolution of about 400~ps,
which is associated with the time jitter of the avalanche photodiode. To
provide a clean measurement of the PL decay, the TR-PL measurements are
conducted at a 3-$\mu$W average pumping power.

\section{Photoluminescence excitation}

PLE experiments allow us to measure the threshold of the optical absorption,
for the direct evaluation of the AZTSe band gap at low temperature. Fig.1
shows the PLE integrated emission intensity for a detection energy window
between 1.05 and 1.09 eV. Knowing that such PLE data do not simply reflect the
absorption spectrum but include also the relaxation processes to the detection
energy, we measured these PLE spectra over various detection energy windows
(not shown here): the spectra are similar whatever the detection energy with
an uncertainty of $\pm3$~meV, showing that the PLE spectra can be, in this
instance, considered to be essentially proportional to the absorption
coefficient \cite{Klingshirn2012_SO}, itself proportional to the density of
states.

\begin{figure}
\includegraphics[width=\columnwidth]{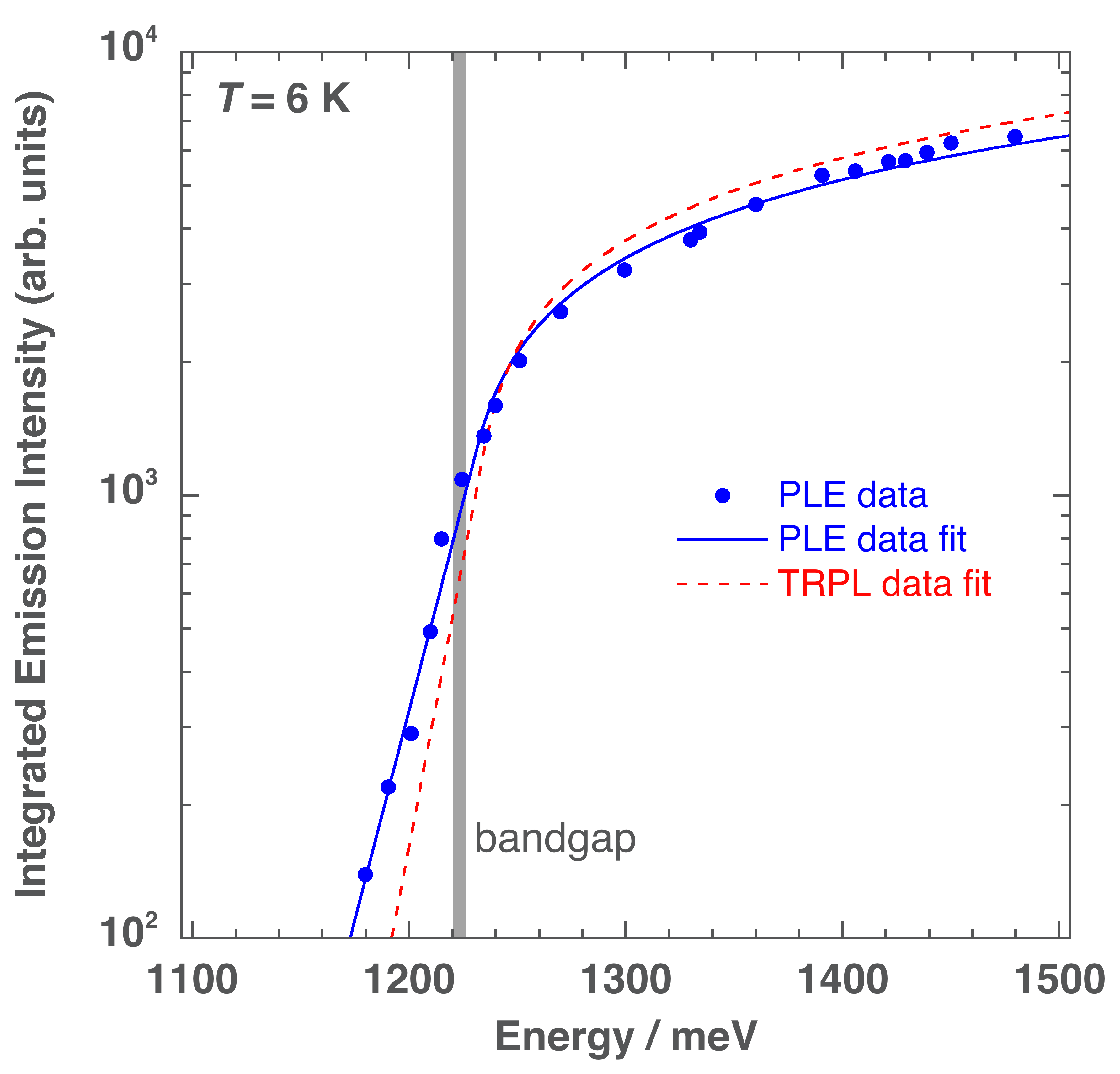}
\caption{\label{fig:PLE} PLE experimental data (dots) at 6~K. The integrated
emission intensity is detected in the 1.05--1.09~eV energy range as a
function of the excitation energy. Least squares fitting of the model density
of states $d(E)$ (solid line --- see text) yields values for the band gap
$\gap = 1220.4$~meV and band tail characteristic energy $U = 23$~meV. The fit
deduced from the TR-PL data is also shown (dashed line).}
\end{figure}

The bandgap is therefore estimated from fitting the PLE data with a square
root density of states in the band, above the band gap, typical of direct band
gap semiconductors, and a density of band tail states in the gap --- related
to the localized states --- described  by a single exponential, as first
proposed by Urbach \cite{Urbach1953_PR92}, which appears to be well adapted to
the case at hand.

More precisely, we model the density of states of such a direct band gap
semiconductor as:
\begin{equation}
\label{density}
d(E)=\begin{cases}
A\,\sqrt{2\,\frac{E-\gap}{U}} & \text{if $E\geq\gap+U/2$} \\
A\,\exp\left(\frac{E-\gap}{U}-\frac{1}{2}\right) & \text{if $E\leq\gap+U/2$}
\end{cases}
\end{equation}
where \gap\, is, as usual, the width of the band gap, and $U$ is the
characteristic energy that scales the extent of the Urbach tail density of
localized defect states in the gap; $A$ is a global amplitude factor. With the
above definition, $d(E)$ and its derivative are both continuous functions at
the connection energy $\gap+U/2$. This is requested by the smooth behavior of
the experimental data around the band gap.

The best least squares fit of the experimental data of figure \ref{fig:PLE}
with this $d(E)$ model function is obtained with the values $\gap =
1220.4$~meV, and $U = 23$~meV. The grey vertical line in figure \ref{fig:PLE}
covers the energy interval between the values taken from the fits of the PLE
data --- 1220.4~meV, and TR-PL data --- 1226.3~meV (see below).

\section{Time-resolved photoluminescence}

Another way to estimate the band gap, and the extent of the band tail of
localized states just below, is to perform a study of TR-PL spectra
\cite{Gokmen2013_APL103,Bleuse2018_JEM47}. These spectra are reported in figure
\ref{fig:TRPL} for various time delays after a 1.55~eV, $\approx200$-fs long,
pulsed excitation\footnote{See Supplemental Material at [URL] for an animated
version of a larger subset (144 spectra) of the experimental data and
corresponding curve fits, from 0.5 to 200~ns}. In figure \ref{fig:TRPL}(a), the
rise of the TR-PL signal is plotted on a linear scale, which shows a maximum
intensity at a 0.90-ns delay: this should correspond to the fast relaxation of
the energy of hot carriers within the bands. Then, in figure \ref{fig:TRPL}(b),
the subsequent signal decay exhibits a red energy shift that increases with
time. This reveals the transfer of carriers within a band tail of states below
the band gap, from the highest states to the lowest ones, on a much longer time
scale --- of the order of 100~ns.

These transfers between states of differing energies explain why, in such a
case, a decay curve measured at a single photon energy can only accidentally
give rise to a single exponential, and would account for only meaningless
characteristic transition times.

In the present case, the time-dependent spectra are plotted, on a
semi-logarithmic scale, for time delays in a geometric series\footnote{Figure
\ref{fig:TRPL}(b) shows 9 spectra for delays from 1 to 200~ns, geometrically
``equidistant'', \emph{i.e.} for delays separated by a factor
$200^{1/8}\approx1.939$, rounded to the time resolution, which is here the
time bin width: 50~ps.}. In such a scheme, if the PL decay were following a
power law dependence on time \cite{Mourad2012_PRB86,Hages2017_AEM7}, the
intensity maxima between two successive curves should be visually equidistant
on this semi-logarithmic plot: it is not the case here, at least for times up
to $\approx 25$~ns, as the larger the time, the closer the spectra look, which
evidences that the PL intensity peak decrease is slower than any power law,
therefore slower than any exponential. For times larger than $\approx 25$~ns,
the data is not so clear, and a power law behaviour cannot be fully excluded.

From this, we can then infer that the carriers at the origin of the emission
in the band tail are not fully independent, since the power law decay inherent
to bimolecular recombination \cite{Mourad2012_PRB86} is absent for delays
below 25~ns. It is also worth mentioning here that, at room temperature, such
a carrier redistribution in CZTSSe was found to occur over only a much
shorter, 1~ns time scale, and on an energy range of only 10~meV
\cite{Hages2017_AEM7}.

To get a more quantitative account of our data, we use the following analysis
methodology: each of these low-resolution, TR-PL spectra is fitted with a
function that is the product of the above density of states $d(E)$ (eq.
\ref{density}) with a Fermi occupation function:
\begin{equation}
f(E) = \left[1+\exp\left(\frac{E-\fermi}{\boltzmann}\right)\right]^{-1}
\end{equation}

Since the band gap \gap\, and the Urbach tail extent $U$ do not depend on
time, the fitting procedure is two-step: the nearly 2000 spectra, from 0.50~ns
to 100~ns after the excitation pulse, are fitted\footnote{A
Levenberg-Marquardt algorithm is used.} with the function $f(E)d(E)$, keeping
the values of \gap\, and $U$ constant, with only 3 parameters varying with
time: $A$, \fermi, and \boltzmann. The cumulated errors between data and
fitted values over these 2000 spectra then represent a global error for the
chosen (\gap, $U$) pair. This is then repeated to find the optimal \gap\ and
$U$ values that minimize this global error \footnote{The minimization search
goes over the rectangle $[1200; 1250]\times[4; 24]$~meV$^2$, where a single
global minimum is found at (1226.3; 16.8)~meV.}.
\begin{figure}
\includegraphics[width=\columnwidth]{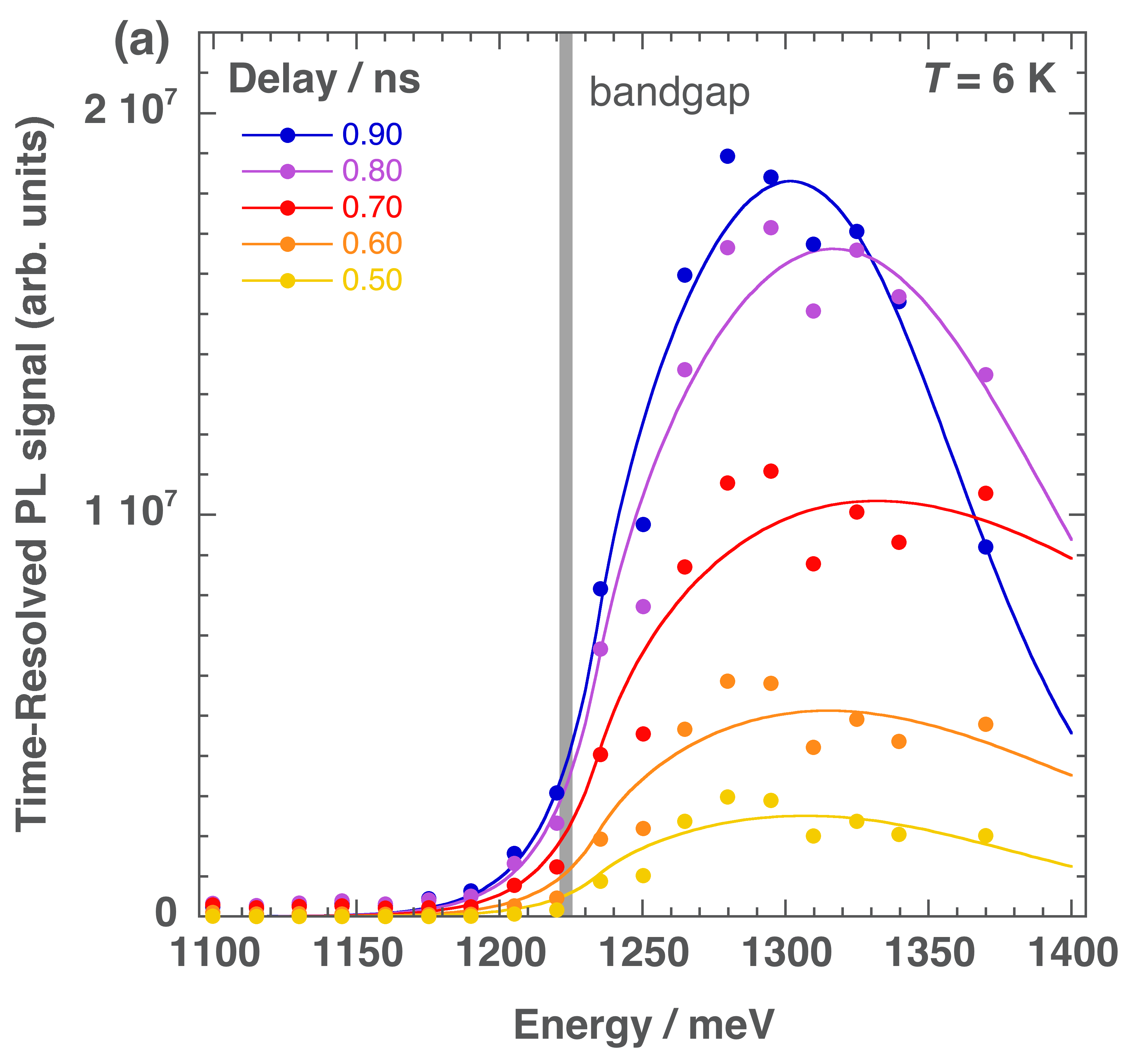}
\includegraphics[width=\columnwidth]{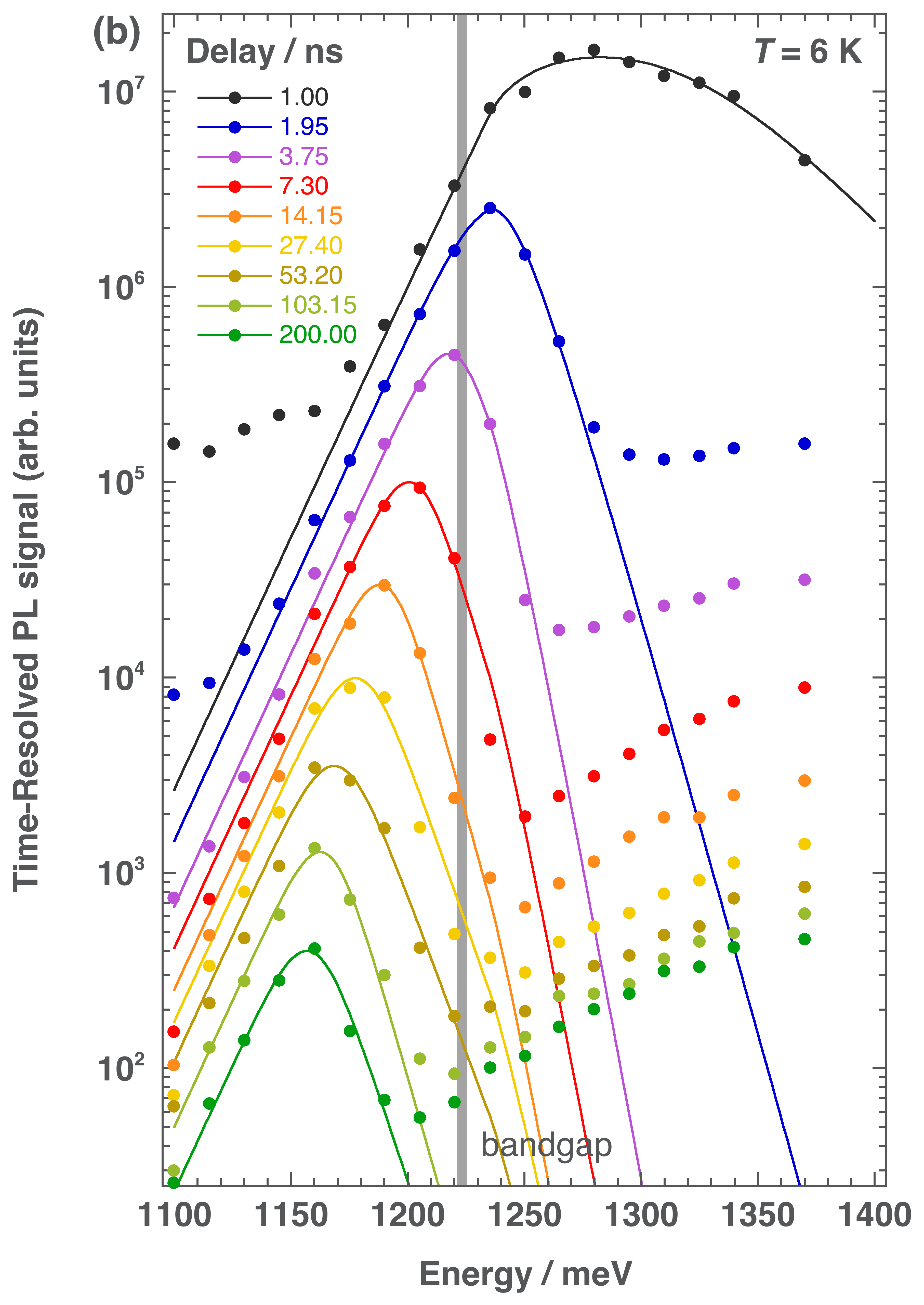}
\caption{\label{fig:TRPL} Low temperature TR-PL spectra. Experimental data
(dots) and fitting curves (solid lines) for various time delays after the
pulse excitation. (a) Signal rise for short delays, on a linear scale. The
TR-PL signal peak is maximum at a 0.90~ns delay. (b) Signal decay, on a
semi-logarithmic scale, for time delays that follow a geometric series from 1
to 200~ns. TRPL spectra exhibit red energy shifts that increase with time,
revealing the transfer of carriers within the band, up to $\approx2$~ns, and
between the localized states in the band tail, for longer delays.}
\end{figure}

The results of this two-step procedure are given in figure \ref{fig:TRPL},
showing the fits (full lines) together with experimental data (points): a
bandgap $\gap = 1226.3\pm3$~meV, and an Urbach tail characteristic energy $U =
16.8\pm1$~meV are precisely found, along with the time variation of the
parameters $A$, \fermi, and \boltzmann. This determination of \gap\, and $U$
is in good agreement with the one deduced above from the PLE experiments. As
in figure \ref{fig:PLE}, the grey vertical line in figure \ref{fig:TRPL}
covers the energy interval from 1220.4 to 1226.3~meV.

\begin{figure}
\includegraphics[width=\columnwidth]{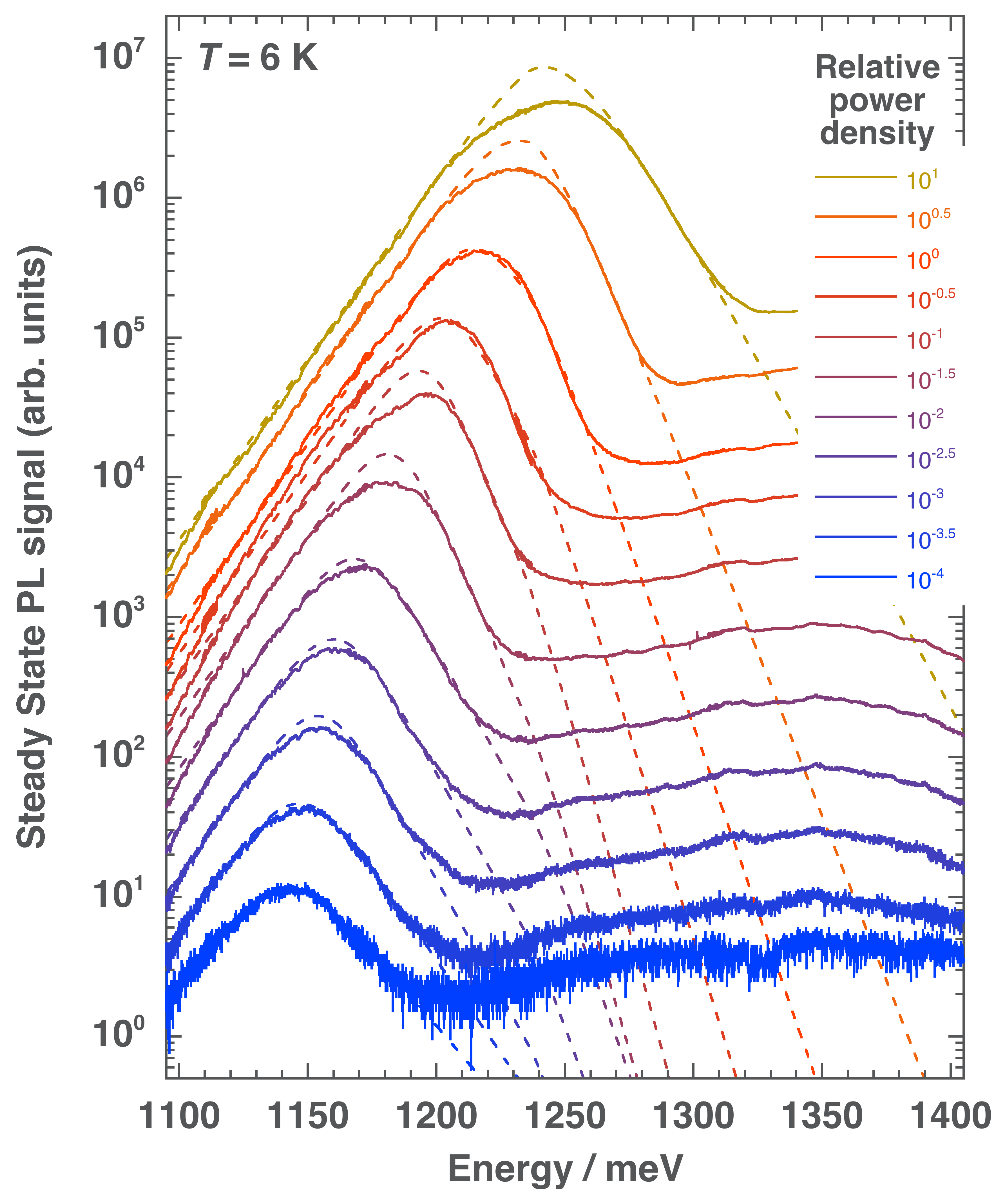}
\caption{Steady-state PL spectra as a function of excitation power --- over 5
orders of magnitude, on a semi-logarithmic scale. The dashed lines
corresponds to fits with the model function $f(E)d(E)$, taking the \gap\, and
$U$ values obtained from the TR-PL data; the \fermi\, parameters extracted
from these fits are reported in figure \ref{fig:peakFWHM}. The exponential
decrease on the low energy side of the spectra correspond to the
characteristic energy $U$ of the Urbach tail density of localized defect
states in the gap.}
\label{fig:PLvsPower}
\end{figure}

One must however take note that we do not currently understand the exact
nature of the emission that is always observed on the high energy side of the
spectra ($E > 1.3$~eV) ; this emission is also observed in steady-state PL
spectra (see figure \ref{fig:PLvsPower}). As it is clearly distinguished from
the emission peak, it is not taken into account in our data analysis, .

\section{Discussion}

It is clear from this analysis that the maxima of the TR-PL peak, for very
short time delays after the pulse, are at energies larger than \gap\, (figure
\ref{fig:TRPL}(a)), and correspond to the recombination of hot carriers,
before and along energy relaxation from phonon emission. On the other hand,
for delays larger than $\approx3$~ns, the maxima of the TR-PL peak occur at
energies lower than \gap, and are related to carrier recombination involving
localized states in the band tail.

\begin{figure}
\includegraphics[width=\columnwidth]{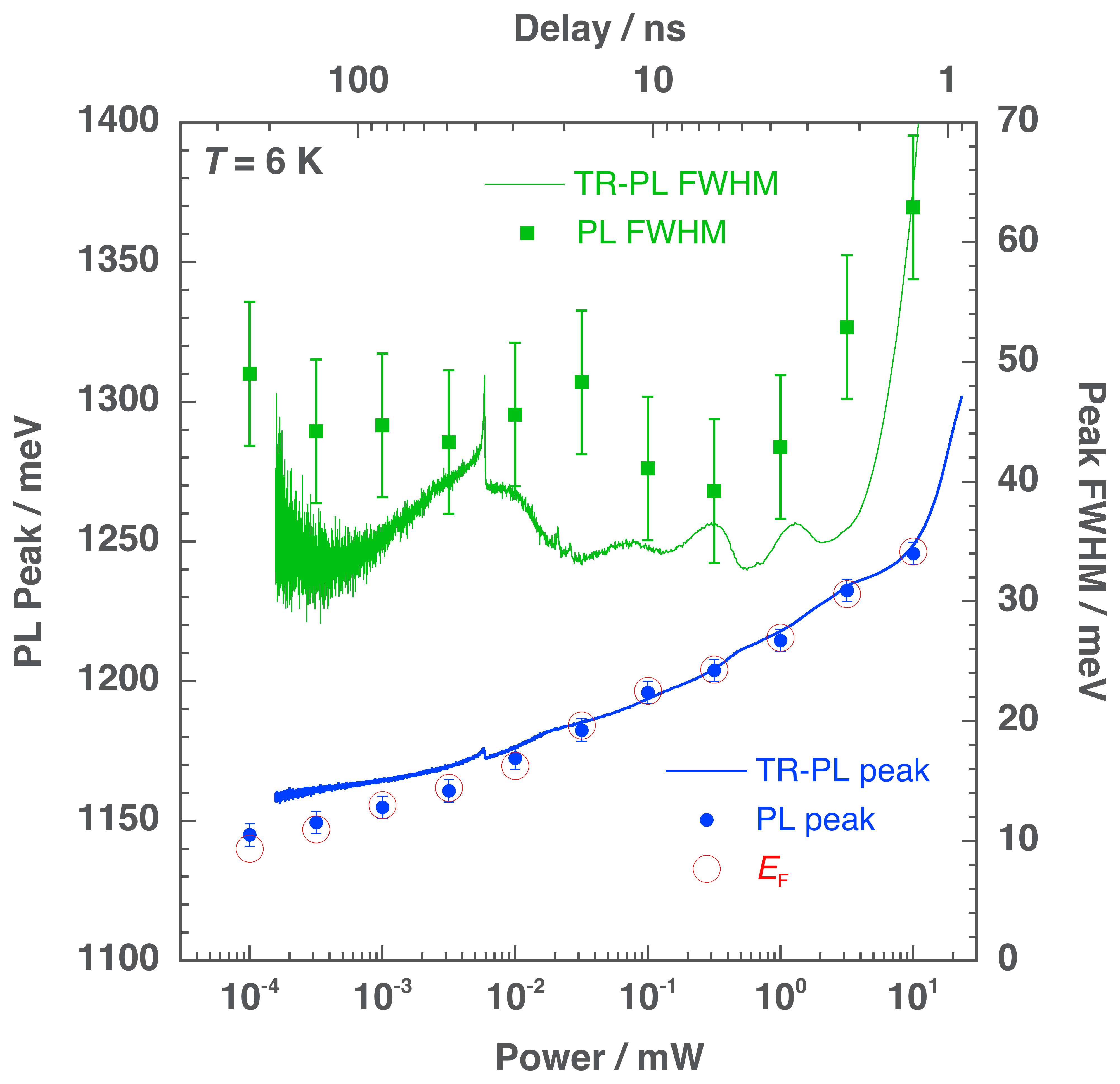}
\caption{Steady-state PL data as a function of excitation power (lower
horizontal, logarithmic axis): the PL peak energy positions (blue dots, left
vertical axis) are closely followed by the values of \fermi\, (open red
circles), which are deduced from the spectral fits in figure 
\ref{fig:PLvsPower}; the experimental values of the FWHM are also reported
(green squares, right vertical axis). Superimposed are TR-PL data as a
function of time delay (upper horizontal, logarithmic axis): the TR-PL peak
energy positions (blue curve, left vertical axis), and peak FWHM (green curve,
right vertical axis) are taken from the fitted curves. The two horizontal,
logarithmic axes give the correspondance between delay and excitation power,
as both correspond to variations of the electron-hole density. The ``jump'' in
TR-PL curves around 36~ns delay comes from a weak, parasitic re-excitation.}
\label{fig:peakFWHM}
\end{figure}

Considering the parameters extracted from the model, the Fermi energy \fermi\,
is --- for delays larger than 1~ns --- a decreasing function of time that
closely follows the peak energy maxima, as expected from the characteristics
of our model function $f(E)d(E)$. This superposition of \fermi\, with the
energy variation of the peak maxima is confirmed with steady-state PL spectra
as a function of power excitation (see  fig.3). There, by varying the power
density over five orders of magnitude, the PL maxima are shifted by
$\approx100$~meV, from 1145 to 1245 meV, as are the values of \fermi\, (see
figure \ref{fig:peakFWHM}).

In both sets of experimental data --- steady-state PL as a function of power
excitation and TR-PL, the emission spectra can be related to the steady-state
or instantaneous carrier density. PL peak and Fermi energies, as well as
carrier temperatures should therefore correspond when the carrier densities
match. This correspondence is attempted in figure \ref{fig:peakFWHM} where we
adjust the excitation power and time scales to show that peak energy and full
width at half maximum (FWHM) pairs exhibit similar trends between steady-state
and time-resolved PL. This shows that, for delays larger than $\approx5$~ns,
the carrier populations in TR-PL measurements are probably close to thermal
equilibrium.

Our second fit parameter, the ``temperature'' \boltzmann, which is directly
related to the peak FWHM, decreases as a function of time and saturates, at
delays larger than 5~ns, to a value of $\approx6$~meV, which is equivalent to
$T\approx70$~K: this temperature, much larger than the bath one (6~K), shows
the limitation of our model to account for weakly-populated discrete defect
states in the band tail. Nevertheless, these values of \boltzmann\, between 1
and 200~ns can be,  used to estimate the FWHM (full green line in figure
\ref{fig:peakFWHM}), which shows values between 35 and 70~meV, fully
accounting for the TR-PL experimental data. These are in very good agreement
with  the peak FWHM deduced from steady-state PL spectra (green squares), and
a factor of 2 smaller than the FWHM usually measured on pure copper
k\"esterites compounds \cite{Bleuse2018_JEM47,Timmo2017_EMRS13}, which lie
between 70 and 110~meV, even for the most ordered CZTS monocrystalline
samples.

\section{Conclusion}

This work reports on a precise determination of the band gap of
monocrystalline \aztse\, at low temperature. Both photoluminescence excitation
and time-resolved photoluminescence lead to a value of \gap\, at 6~K equal to
$1223\pm3$~meV. It is worth noting that such values are much lower than the
ones reported at room temperature for polycristalline thin films: figuring
diffuse reflectance data in a Tauc plot, W. Gong et al \cite{Gong2015_PSSc12}
obtained a value of $\gap = 1.34$~eV, whereas Gershon et al.
\cite{Gershon2016_AEM6_22} deduced a value of 1.35~eV from External Quantum
Efficiency data, in agreement with their room-temperature PL emission maximum.
Variations of the gap values can however be strongly dependent on the
selenization temperature, \emph{i.e.} Se content, as pointed out by Jiang et
al. \cite{Jiang2019_JAP125} who reported band gap values between 1.33~eV and
1.59~eV. Our results nevertheless indicate that the usual behavior of
semiconductor band gap --- a decrease with temperature because of
electron-phonon coupling, does not hold here: further studies as a function
of temperature are under way to elucidate this unexpected behavior.

This work also presents a quantitative methodology to precisely deduce the
band gap \gap\, and the characteristic energy $U$ for the extent of the tail
of localized states, from the PLE and TR-PL data. Such an approach is
different from the one usually used, namely measuring the energy shift between
the PL emission and the absorption PLE threshold (the Stokes shift). The
advantage of the present method is that no arbitrary choice of the low power
excitation has to be done to select the PL emission spectrum and its peak
energy (see figure \ref{fig:peakFWHM}).

Finally, with this systematic optical spectroscopy study and the above
analysis methodology, the characteristic energy of the band tail in this new
\aztse\, material is found to be $20\pm3$~meV, which leads to a
photoluminescence FWHM that is half those usually reported for even the
most-ordered Cu-based k\"esterites compounds.

\begin{acknowledgments}
This research was supported by the french ANR program ``Carnot Energies du
Futur'', under the project name CAZTS.
\end{acknowledgments}

\bibliography{AZTSe_bandGapAndTail}

\end{document}